\documentclass{cas-sc}

\usepackage{natbib}
\setcitestyle{numbers,square}
\usepackage{subfigure}
\usepackage{float}
\usepackage{graphicx}
\usepackage{tabularx}
\usepackage{booktabs}
\usepackage[labelfont=bf,format=plain,justification=raggedright,singlelinecheck=false]{caption}

\def\tsc#1{\csdef{#1}{\textsc{\lowercase{#1}}\xspace}}
\tsc{WGM}
\tsc{QE}
\tsc{EP}
\tsc{PMS}
\tsc{BEC}
\tsc{DE}

\begin{document}

% Short author
\shortauthors{Xiangxi Li et~al.}

\title [mode = title]{Effect of antibody levels on the spread of disease in multiple infections}               
\author[1]{Xiangxi Li}

\affiliation[1]{organization={College of Artificial Intelligence, Southwest University},
    city={Chongqing},
    postcode={400715}, 
    country={China}}

\author[2]{Yuhan Li}
\affiliation[2]{organization={School of Mathematical Sciences, Queen Mary University of London},
    city={London},
    postcode={E1 4NS}, 
    country={United Kindom}}

\author[1]{Minyu Feng}
\fnmark[*]

\ead{myfeng@swu.edu.cn}
\cortext[cor1]{Corresponding author}

\author[3,4]{Jürgen Kurths}
\affiliation[3]{organization={Potsdam Institute for Climate Impact Research},
    city={Potsdam},
    postcode={14437}, 
    country={Germany}}

\affiliation[4]{organization={Department of Physics, Humboldt University},
    city={Berlin},
    postcode={12489}, 
    country={Germany}}

\begin{abstract}
There are complex interactions between antibody levels and epidemic propagation; the antibody level of an individual influences the probability of infection, and the spread of the virus influences the antibody level of each individual. There exist some viruses that, in their natural state, cause antibody levels in an infected individual to gradually decay. When these antibody levels decay to a certain point, the individual can be reinfected, such as with COVID-19. To describe their interaction, we introduce a novel mathematical model that incorporates the presence of an antibody retention rate to investigate the infection patterns of individuals who survive multiple infections. The model is composed of a system of stochastic differential equations (SDE) to derive the equilibrium point and threshold of the model and presents rich experimental results of numerical simulations to further elucidate the propagation properties of the model. We find that the antibody decay rate strongly affects the propagation process, and also that different network structures have different sensitivities to the antibody decay rate, and that changes in the antibody decay rate cause stronger changes in the propagation process in Barabási–Albert (BA) networks. Furthermore, we investigate the stationary distribution of the number of infection states and the final antibody levels, and find that they both satisfy the normal distribution, but the standard deviation is small in the Barabási–Albert (BA) network. Finally, we explore the effect of individual antibody differences and decay rates on the final population antibody levels, and uncover that individual antibody differences do not affect the final mean antibody levels. The study offers valuable insights for epidemic prevention and control in practical applications.

\end{abstract}

% \begin{highlights}
% \item The article introduces a novel mathematical model that incorporates the presence of an antibody retention rate to investigate the infection patterns of individuals who have survived multiple infections.

% \item The model uses a system of stochastic differential equations (SDE) to derive the equilibrium point and threshold of the model and presents rich experimental results of numerical simulations to further elucidate the propagation properties of the model.
% \item 
% The study finds that the antibody decay rate strongly affects the propagation process, and also that different network structures have different sensitivities to the antibody decay rate.
% \end{highlights}


\maketitle

\section{Introduction}
With the rapid advancements in information technology, the field of complex networks has made remarkable progress by integrating multiple disciplines. Complex networks possess a unique structure that facilitates nodes to exhibit a high degree of connectedness and interaction, leading to intricate network structures and behaviors. Moreover, the characteristics of complex networks provide an ideal environment for various types of transmission. The establishment of mathematical foundations for transmission models by Kermack and McKendrick \cite{kermack1927contribution} has led to the development of various mathematical methods, substantially improving our understanding of epidemic spreading. In recent times, there has been an increasing number of articles utilizing physics methods to study various social phenomena. Jusup et al. have systematically reviewed the application of these methods \cite{jusup2022social}. Consequently, different propagation models have been well applied on complex networks in recent years \cite{liang2020mathematical,wang2019coevolution,wang2019impact,xia2019new,kabir2019analysis}. The outbreak of COVID-19 has brought infectious disease control to the forefront, and the study of transmission models can offer a theoretical support for controlling the spread of epidemics  \cite{hancean2022occupations,ul2021modeling,sun2020did,he2020seir,small2020modelling,thurner2020network,ribeiro2020city}. Therefore, the application of infectious disease models on complex networks holds significant value.

The propagation on complex networks originated from network modeling. In recent years, Perc et al. provided a brief overview of the diffusion dynamics and information spreading \cite{perc2019diffusion}. Feng et al. have used the birth-death process to model and analyze networks in various ways \cite{feng2022heritable,li2023evolving,zeng2023temporal}. On the basis of complex network modeling, scholars have extensively studied various factors influencing the transmission of infectious diseases, as documented in numerous studies including those by Xie et al. \cite{xie2023contact}, Li et al. \cite{li2022limited,li2022network}, Connolly et al. \cite{connolly2021extended}, and Hernandez et al. \cite{hernandez2020evaluating}. One critical aspect that has garnered recent attention is the role of antibody production following a COVID-19 infection. Xia et al. reported that antibody levels reached a plateau 16-30 days after symptom onset, gradually declining to a steady state after about four months \cite{xia2021longitudinal}. Separate research by Yin et al. \cite{yin2022impact}. introduced a three-layer coupled network model to explore the complex interplay between negative vaccine-related information, vaccination behavior, and epidemic spread.
Liu et al. investigated the interaction between epidemic spreading and awareness diffusion in a two-layer network model. They also explored the impact of individual heterogeneity on the epidemic threshold. Their findings suggest that by promoting more effective information dissemination and enhancing group interactions within the awareness layer, the spread of the epidemic can be significantly suppressed \cite{liu2023epidemic}. Jardón-Kojakhmetov et al. analyzed fast-slow versions of epidemiological models such as SIR, SIRS, and SIRWS, with the SIRWS model being particularly relevant due to its inclusion of a W-zone for populations with declining immunity \cite{jardon2021geometric}.
Leung et al. developed a dual-pathogen transmission model to investigate how immune enhancement and cross-immunity influence the timing and severity of epidemic transmission, a topic of significant importance given the potential for COVID-19 reinfection \cite{leung2016periodic}. They further proposed a model distinguishing between primary and secondary infections to better understand the interaction between infection and immunity \cite{leung2018infection}.
In the traditional SIRWS model, it has been commonly assumed that the rate of transition from the immune state (R) to the waning state (W), and from the waning state (W) back to the susceptible state (S), is uniform. However, Opoku-Sarkodie et al. relaxed this assumption by allowing for an asymmetric division of the entire immunity period, highlighting that the duration of the waning period is a crucial parameter affecting long-term epidemiological dynamics \cite{opoku2022dynamics}.
Apio et al. emphasized the significance of antibody studies and provided COVID-19 antibody rates with 95\% confidence intervals for the Korean population, based on recent antibody tests conducted in Korea \cite{apio2020confidence}. Wang et al., through a networked metapopulation model, analyzed the effects of migration on the spread of epidemics \cite{wang2022epidemic}.
These studies collectively enhance our understanding of the intricate dynamics of infectious disease transmission and the role of various factors, including antibody production, vaccination behavior, and network structures, in shaping these dynamics.

In the study of disease transmission on complex networks, statistical methods are frequently employed to examine a variety of properties. Deng et al. utilized gamma, Weibull, and lognormal distributions to estimate the incubation period of diseases \cite{deng2021estimation}. The mathematical modeling of epidemics provides valuable insights, such as predicting the size of an epidemic and determining the critical intervention level for effective disease control, as noted by Grassly et al \cite{grassly2008mathematical}.
Pastor-Satorras et al. reviewed different network distributions, including the degree distribution and the distribution of the number of infections, which are essential in understanding the spread of diseases \cite{pastor2015epidemic}. Feng et al. introduced evolving network models that incorporate birth and death processes, akin to queuing systems in mathematics, to account for both the growth and decline of network vertices. They further investigated how individuals with varying properties influence the spread of diseases \cite{feng2018evolving} \cite{feng2023impact}.
Gosak et al. applied a stochastic model to various social networks to evaluate the impact of community lockdowns and travel restrictions on epidemic control \cite{gosak2021community}. Li et al. used an open Markov queueing network model to study the distribution of individuals across different epidemic states and presented a model of an evolving population network that considers the migration of individuals \cite{li2021protection}. Fan et al. studied the dynamic spread of epidemics on multilayer networks that include degenerate complexes. They found that considering higher-order interactions, where connections may involve more than two individuals, significantly impacts the epidemic threshold and spread \cite{fan2022epidemics}.

However, the existing literature predominantly focuses on scenarios where a virus infects an individual once or extends the classical SIRS model by introducing a W state. Building upon the prior studies, we present and analyze a novel mathematical model that incorporates the presence of an antibody retention rate to investigate the infection patterns of individuals who have survived multiple infections. In contrast to previous models \cite{strube2021role,calvetti2023post}, we set the antibody of each individual as a continuous variable, and employ a stochastic process approach to represent the variation of individual antibody levels. Moreover, the probability of infection corresponding to different antibody levels is also a continuously variable, which more accurately reflects the real-world situation. The principal contributions of this study are threefold. Firstly, we introduce the antibody retention rate into the existing SIRS model and explicitly depict the process of transformation from the R to the S state, along with the process of infection for individuals in the I state. Secondly, we utilize a system of stochastic differential equations to derive the equilibrium point and threshold of the model. Finally, we present rich experimental results of numerical simulations to further elucidate the propagation properties of the model. Based on our findings, we can offer valuable insights for epidemic prevention and control in practical applications.

This paper is organized as follows: In Sec. \uppercase\expandafter{\romannumeral2}, we introduce the process of model building. In Sec. \uppercase\expandafter{\romannumeral3}, we perform numerical simulations. Finally, discussions, conclusions, and outlooks are given in Sec. \uppercase\expandafter{\romannumeral4}.

\section{Model descriptions}

The study of mathematical models for infectious disease transmission has become a crucial area of research in the field of epidemiology. These methods provide a better understanding of the structure of realistic human connections and social networks, which allow for more accurate descriptions and predictions of disease transmission dynamics. Here, our study delve into the significance of antibodies in the spread of epidemics. Antibodies are crucial defense mechanisms employed by organisms to combat pathogens. They perform this function through a variety of means. By developing a complex network model, we can gain a more comprehensive understanding of the role and mechanisms of antibodies in virus transmission from a macroscopic standpoint. This understanding enables us to more accurately predict the impact of antibodies on epidemic transmission, as well as the impact of vaccination on epidemic control. 

In this section, we introduce an SIRS model that accounts for varying antibody levels in each individual to describe the potential of multiple infections, as observed in COVID-19. We subsequently formulate a system of stochastic differential equations to investigate the equilibrium point and threshold of the model.

The model uses a lot of symbols, and Table 1 provides a detailed explanation of the meanings of each symbol.

\begin{table}[h]
	\caption{Symbols explanation}
	\begin{tabularx}{\textwidth}{p{0.08\textwidth}X}
		\toprule	
		\underline{\textbf{\emph{Parameters}}} \\  
		$\beta$ & the basic infection rate \\
		$\mu$      & the recovery rate \\
		$\gamma$    & the rate that R-state individuals return to the S state \\
		$\alpha$     & the average of antibody levels (in OU process)   \\
		$\theta$     & the rate coefficient of regression to the mean value (in OU process)  \\
		$\eta$   & the amount of antibodies acquired by the individual after infection\\
		$\sigma$   & the intensity of the noise\\
            $\psi$   & the mean antibody level of the population at steady state\\
		$N$   & the total number of individuals in the network\\
		$\alpha_p$ & a slope parameter that controls the degree to which the antibody level affects the probability of infection\\
		$\gamma_p$   &  a threshold parameter that controls the inflection point of the impact of the antibody level on the probability of infection\\
		$\mu_{Gauss}$ &  the mean of the normal distribution\\
		$\sigma_{Gauss}$   & the standard deviation of the normal distribution\\
            
		\underline{\textbf{\emph{Random Variables}}} \\ 
		$A_{i}(t)$  & the antibody level of individual $i$ at time $t$\\
		$P_{infect}(i)$   & the probability of infection for individual $i$  \\
		$S_{i}(t)$   & the susceptible state of the $i_{th}$ individual at time $t$  \\
		$I_{i}(t)$ & the infected state of the $i_{th}$ individual at time $t$  \\
		$R_{i}(t)$  & the recovered state of the $i_{th}$ individual at time $t$ \\
		
		\bottomrule
	\end{tabularx}
\end{table}

\subsection{SIRS Model with Antibody Levels}
Consider a network comprising $N$ individuals, where each individual can exist in one of three states: susceptible ($S$), infected ($I$), or recovered ($R$). We represent the susceptible state of the $i_{th}$ individual at time $t$ as $S_i(t)$, the infected state as $I_i(t)$, and the recovered state as $R_i(t)$. By adopting the SIRS model, the state transition of each individual can be characterized as
\begin{equation}
    \left\{
    \begin{aligned}
    &d S_{i}(t) =\left[-\beta S_{i}(t) \sum_{j=1}^{N} B_{i j} I_{j}(t)+\gamma R_{i}(t)\right] d t \\
    &d I_{i}(t) =\left[\beta S_{i}(t) \sum_{j=1}^{N} B_{i j} I_{j}(t)-\mu I_{i}(t)\right] d t\\
    &d R_{i}(t) =\left[\mu I_{i}(t)-\gamma R_{i}(t)\right] d t
    \end{aligned}
    \right.
\end{equation}
Herein, $\beta$ represents the basic infection rate, which signifies the average number of susceptibles that an infected individual is expected to transmit the infection to per unit time. Similarly, $\mu$ denotes the recovery rate, representing the average proportion of infected individuals that will recover per unit time, $\gamma$ is the rate that R-state individuals return to the S state. The adjacency matrix $B$ is defined such that $B_{ij}$ reflects whether there exist edges between individuals $i$ and $j$ \cite{harary1962determinant}.

In assessing the impact of antibodies, it is posited that each individual possesses a quantifiable level of antibodies, denoted as $A_i(t)$ for individual $i$. In the spread of an epidemic, an individual's antibody levels may rise due to infection, but over time, if there is no re-exposure to the virus, the antibody levels will decrease due to natural decay, a process that can be described by the Ornstein-Uhlenbeck (OU) process. An important characteristic of the OU process is its "mean-reverting" nature, meaning that the variable fluctuates around some long-term average. This corresponds to the phenomenon where antibody levels gradually rise after infection and then gradually decline over time, tending towards a baseline level. The varations in antibody levels across individuals exhibit analogous characteristics. Under normal conditions we set the average value of the antibody level to 0. This will cause the antibody level to gradually decrease after the individual is infected, and the closer to 0 the slower the rate of antibody decrease. The expression of antibody level $A_i(t)$ is shown in Eq. 2.
\begin{equation}
    d A_{i}(t)=\theta\left(\alpha-A_{i}(t)\right) d t  +\sigma d W_{i}(t)
\end{equation}
where $\theta$ denotes the rate coefficient of regression to the mean value, while $\alpha$ represents the mean value, signifying the average of antibody levels. $W_i(t)$ corresponds to the Brownian motion, representing the random noise. Additionally, $\sigma$ reflects the intensity of the noise, which is equivalent to the standard deviation of the Brownian motion.

The incorporation of random variations in individual antibody levels is essential to appropriately model the immune system of individuals. As immune systems can vary even in the same environment, the use of the OU process provides a suitable means of accounting for this stochasticity, thereby enabling a more accurate description of the differences and variations between individuals. Moreover, the Brownian motion in the OU process characterizes the stochastic perturbation of antibodies by the distinct behaviors of each individual, thereby more precisely capturing the changes in antibody levels over time. The overall process can be depicted as Fig. 1.

\begin{figure}[h]
    \centering
    \includegraphics[scale = 0.45]{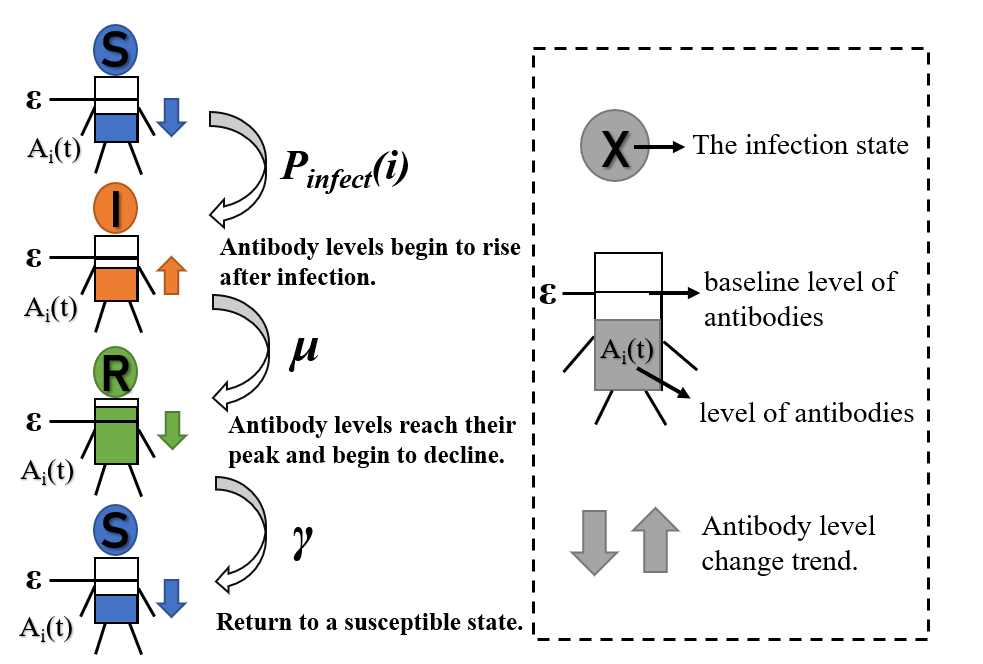}
    \caption{\textbf{The propagation process.} Where the left side is filled to represent the individual's antibody level. The turning point of the antibody level is represented by $\epsilon$, this value also represents a general antibody level regression value, indicating the state of an individual's antibody level after infection. The middle section of the diagram describes the probability and conditions of transmission, and the right provides a detailed explanation of the significance of each panel.}
    \label{fig1}
\end{figure}

In the case that the probability of infection for each individual depends on their level of antibodies, where a higher antibody level leads to a lower probability of infection, we introduce a new function. It represents the probability of infection for individual $i$, with $P_{infect}(i)$ being a monotonically decreasing function of the antibody level $A_i(t)$. In doing so, the model can better reflect the impact of antibody levels on the probability of infection, as individuals with elevated antibody levels tend to exhibit a reduced probability of contracting the infection. To illustrate this connection, we can mathematically describe the relationship using a sigmoid function in Eq. 3, which effectively portrays the gradual transition from a higher probability of infection for individuals with lower antibody levels to a lower probability for those with higher levels of antibodies.

\begin{equation}
    P_{infect}(i) = \beta / (1 + e^{(\alpha_p (A_i(t) - \gamma_p))})
\end{equation}
where $P_{infect}(i)$ is the probability of individual $i$ becoming infected, $\beta$ is the basic probability of infection same as in Eq. 1, $\alpha_p$ is a slope parameter that controls the degree to which the antibody level affects the probability of infection, and $\gamma_p$ is a threshold parameter that controls the inflection point of the impact of the antibody level on the probability of infection. The adoption of a nonlinear function ensures that the probability of infection is extremely low when the antibody level is high and only suddenly increases when the antibody level drops to a certain level, and the function curve of this function is shown in Fig. 3. It's important to note that the variable $\beta$ here doesn't directly represent the infection probability; rather, the infection probability is denoted by $P_{infect}(i)$. $\beta$ serves as a parameter influencing the infection probability $P_{infect}(i)$. For instance, when an individual's antibody level $\alpha$ is at 0, their infection probability would be $\beta / (1 + e^{-(\alpha_p \gamma_p)})$.

Based on the above description, we extend the SDE of the standard SIRS model as follows.
\begin{equation}
    \left\{
    \begin{aligned}
    &d S_{i}(t) =-P_{infect}(i) S_{i}(t) \sum_{j=1}^{N} B_{i j} I_{j}(t) d t+\gamma R_{i}(t) d t \\
    &d I_{i}(t) =P_{infect}(i) S_{i}(t) \sum_{j=1}^{N} B_{i j} I_{j}(t)dt-\mu I_{i}(t) d t\\
    &d R_{i}(t) =\mu I_{i}(t)dt-\gamma R_{i}(t) d t \\
    &d A_{i}(t)=\theta\left(\alpha-A_{i}(t)\right) d t +\sigma W_{i}(t) d t
    \end{aligned}
    \right.
\end{equation}

Eqs. 3 and 4 outline a comprehensive propagation process. In contrast to the conventional SIRS model that disregards the influence of antibodies, our model incorporates the notion of antibodies to depict the immune status of individuals with greater precision. Additionally, we establish a correlation between an individual's antibody levels and their probability of infection, thus increasing the applicability of the model in real-life scenarios.

In order to analyze the equilibrium point of the model, We need to find a steady-state solution of the system where \(\frac{dS}{dt} = \frac{dI}{dt} = \frac{dR}{dt} = 0\). This means that in the steady state, the number of susceptibles, infectives, and recovereds does not change over time. Alternatively, in the steady state, while the antibody levels of individual organisms may fluctuate due to infections, the average antibody level of the population will remain around a certain value. We assume this value to be $\psi$, and we use the average degree $\langle k\rangle$ instead of $\sum_{j=1}^{N} B_{i j}$, $N$ represents the total number of individuals in the network. $\langle k\rangle$ represents the average degree of the network, which indicates the average number of edges that each node has with its neighboring nodes, and there is no dynamical and structural (related to network) correlations and the individuals dependence are replaced to the average value of such quantities. Therefore, the index $i$ in $P_{infect}(i),  S_{i}, I_{i}$ are replaced by $S, R, I$ and $P_{infect}$ indicating these
quantities do not depend on the individual $i$. Then we can express:

\begin{equation}
    \left\{
    \begin{aligned}
    & -P_{infect} S I \langle k\rangle+\gamma R = 0 \\
    & P_{infect} SI \langle k\rangle-\mu I = 0 \\
    & \mu I-\gamma R = 0 \\
    & P_{infect} = \beta / (1 + e^{(\alpha_p (\psi - \gamma_p))})
    \end{aligned}
    \right.
\end{equation}

Additionally, since the total population size remains constant, we can derive:

\begin{equation}
    \begin{aligned}
    N &=S+I+R  \\
    \end{aligned}
\end{equation}

Combining Eq. 5 and Eq. 6, we can solve for:

\begin{equation}
    \left\{
    \begin{aligned}
    &S =\frac{\mu }{P_{infect} \langle k\rangle }   \\
    &I = \frac{N\mu \gamma P_{infect}  \langle k\rangle - \mu^2 \gamma}{P_{infect}  \langle k\rangle \mu (\mu+\gamma) }   \\
    &R = \frac{N\mu P_{infect}  \langle k\rangle - \mu^2}{P_{infect}  \langle k\rangle(\mu+\gamma) }   \\
    & P_{infect} = \beta / (1 + e^{(\alpha_p (\psi - \gamma_p))}) \\
    \end{aligned}
    \right.
\end{equation}

 Eq. 7 represents the relationship between the expected value of each state variable $(S, I, R)$ and the probability of infection. This equation describes how the expected number of susceptible $(S)$, infected $(I)$, and recovered $(R)$ individuals in a population changes based on the probability of infection.

\subsection{Threshold analysis}
In this section, we will discuss the theoretical threshold of the model. The threshold refers to a critical parameter value, in this case, it is \(\beta\). When \(\beta\) exceeds this value, the disease can persist and lead to a large-scale epidemic within the network. This threshold defines the critical point between the disease's extinction and its transition to a widespread epidemic.

When the disease propagation is near the threshold $(A_i, I_i\to0)$, there exists a set of locally stable solutions to system Eq. (4) such that near the equilibrium point the system converges to $(S,I,R) = (1,0,0)$ , in which state there are almost no cases of individuals contracting the disease twice in a row, and $S, I, R$ represents the proportion of individuals in the corresponding states. So we assume that the steady antibody level $A_i (t) = \psi$, from which we obtain Eq. 8.
\begin{equation}
    P_{infect}(i) = \beta / (1 + e^{\alpha_p (\psi - \gamma_p)})
\end{equation}

Simultaneously, we assume that the network is homogeneous. According to the normalization condition $S_i + I_i + R_i = 1$, we rewrite the system Eq. 4 as follows:

\begin{equation}
    \left\{
    \begin{aligned}
    &\frac{d S}{dt} =-\frac{\beta}{(1 + e^{\alpha_p (\psi - \gamma_p)})} \langle k\rangle S  I+\gamma (1 -S -I) \\
    &\frac{d I}{dt} =\frac{\beta}{(1 + e^{\alpha_p (\psi - \gamma_p)})} \langle k\rangle S  I-\mu I\\
    \end{aligned}
    \right.
\end{equation}

At the equilibrium point there is $\frac{d S}{dt}=\frac{d I}{dt} = 0$, therefore the Jacobian matrix of Eq. 9 at the equilibrium point $(S,I,R) = (1,0,0)$ is

\begin{equation}
    \boldsymbol{J}=\left[\begin{array}{cc}
    -\gamma & -\frac{\beta}{(1 + e^{\alpha_p (\psi - \gamma_p)})} \langle k\rangle-\gamma \\
    0, & \frac{\beta}{(1 + e^{\alpha_p (\psi - \gamma_p)})} \langle k\rangle-\mu
    \end{array}\right]
\end{equation}

To ensure stability of the system Eq. 4 near the equilibrium point, it is necessary for the eigenvalues of $\boldsymbol{J}$ to satisfy the condition of negativity, i.e., $-\gamma < 0$ and $\frac{\beta}{(1 + e^{\alpha_p (\psi - \gamma_p)}} \langle k\rangle - \mu < 0$. Based on these conditions, it can be deduced that:

\begin{equation}
    \beta < \frac{\mu(1 + e^{\alpha_p (\psi - \gamma_p)})}{\langle k\rangle} 
\end{equation}

As can be seen from Eq. 11, the transmission threshold is related to the disease recovery rate $\mu$ and the average degree $\langle k\rangle$, and is also influenced by the parameters of the infection probability function $P_{infect}(i)$. When the disease recovery rate is larger, the threshold of transmission increases with it, while when the average degree of the network becomes larger, the threshold of transmission decreases.

\section{Numerical simulation}
In this section, we focus on studying the propagation process through numerical simulation experiments. Our simulation experiments were conducted in a Python 3 environment, with the network construction parameters set as a total of $N = 1000$ individuals, average degree $k = 4$, and the edge reconnection probability $p = 0.1$ for the WS network, The disease's basic transmission rate $\beta$ is 0.2, the recovery rate $\mu$ is 0.1, and the initial average antibody level $\alpha$ = 0. $\sigma$ equals 0.01, and $\theta$ is 0.001. Firstly, we will visually present the two new equations that we have defined in this article. Then, we will demonstrate how our parameters affect these two variables. We will run our model on a simulated network and record the changes in the relevant variables under different parameters. In order to make the simulation more meaningful, we used a variety of different antibody decay rates to mimic real-life scenarios. We present a graphical representation of the two newly defined equations, $A_i(t)$ and $P_{infect}(i)$.
\begin{figure}[h]
    \centering
    \subfigcapskip=2pt
    \subfigure[\textbf{$A_i(t)-\sigma-\theta$}]{
        \includegraphics[width = 0.45\linewidth]{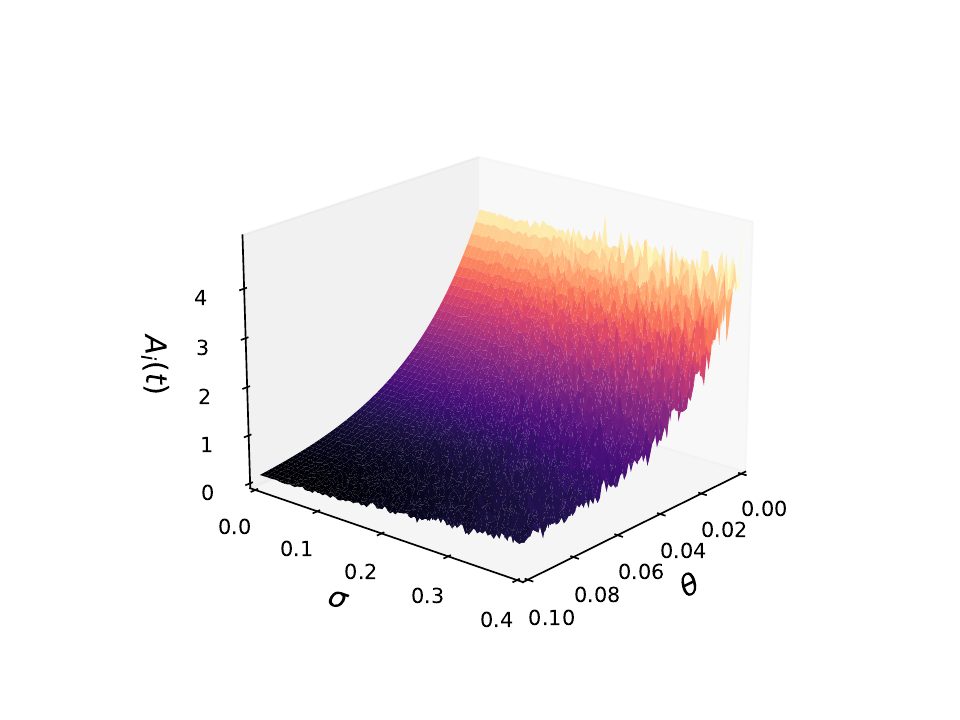}}
    \subfigure[\textbf{$P_{infect}(i)-\gamma_p-\alpha_p$}]{
        \includegraphics[width = 0.45\linewidth]{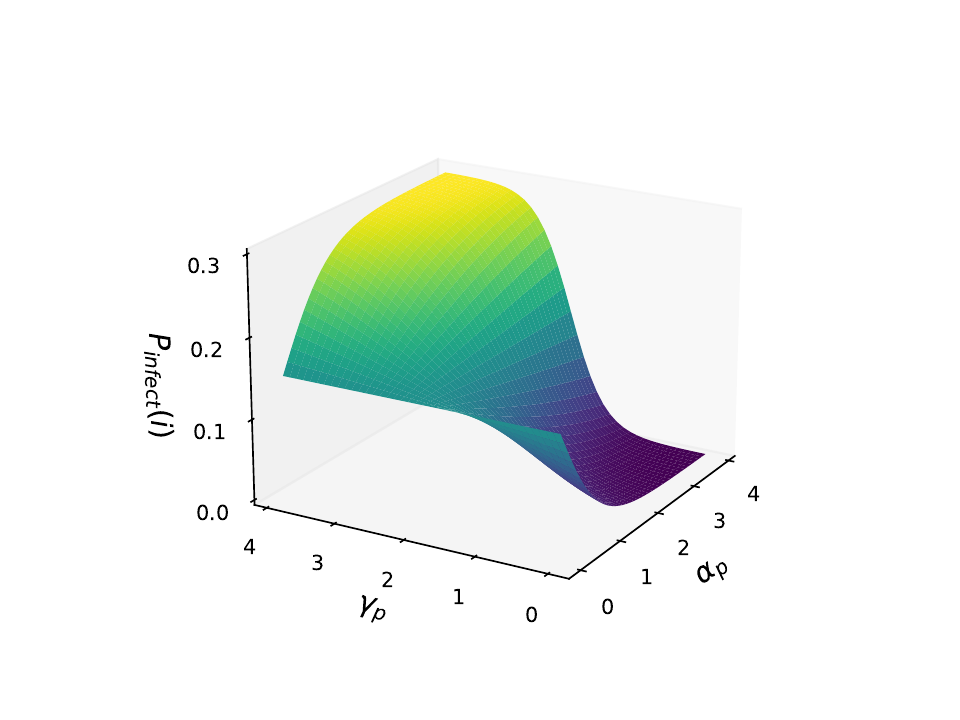}}
    \caption{\textbf{Visual representation.} Figure (a) displays the WS network with $N = 1000$, $k = 4$, and $p = 0.1$, along with the basic infection rate $\beta = 0.2$, disease recovery rate $\mu = 0.1$, and antibody level reversion value $\alpha = 0$, $\alpha_p = 3$, $\gamma_p = 1.2$. By varying the values of $\theta$ and $\sigma$ and allowing the propagation to continue for 50 time steps, we obtain the mean level of antibodies in the network. Figure (b) illustrates the changes in the infection probability $P_{infect}(i)$ with respect to the logarithm of $\gamma_p$ and $\alpha_p$.}
\end{figure}

\begin{figure}[h]
    \centering
    \subfigcapskip=2pt
    \subfigure[\textbf{Effects of $\alpha_p$ on $P_{infect}(i)$ in Eq. 8}]{
        \includegraphics[width = 0.39\linewidth]{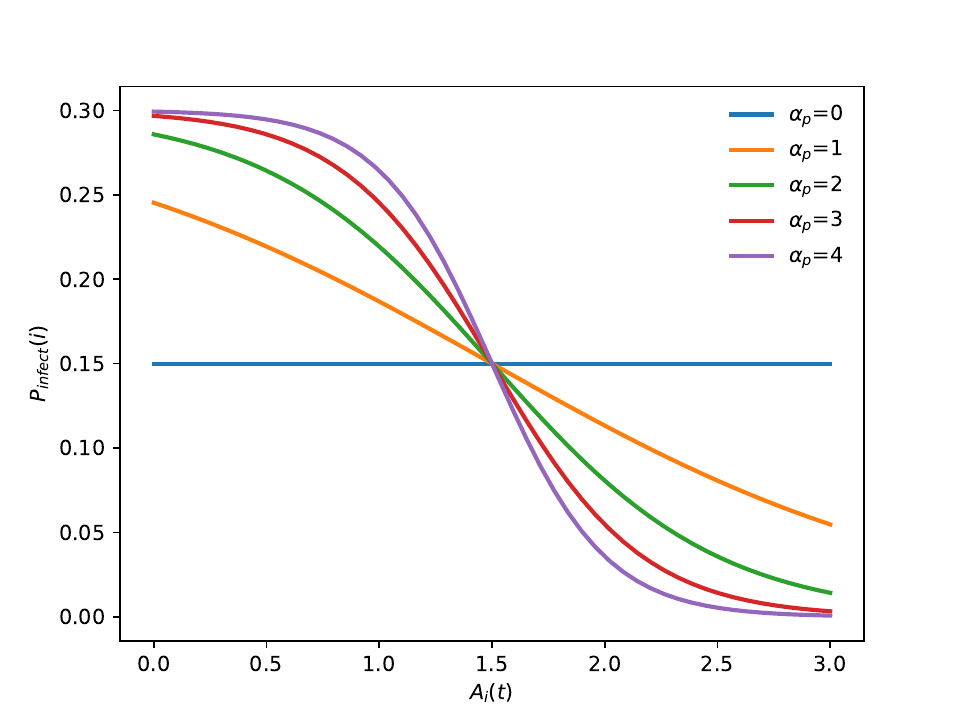}}
    \subfigure[\textbf{Effects of $\gamma_p$ on $P_{infect}(i)$ in Eq. 8}]{
        \includegraphics[width = 0.39\linewidth]{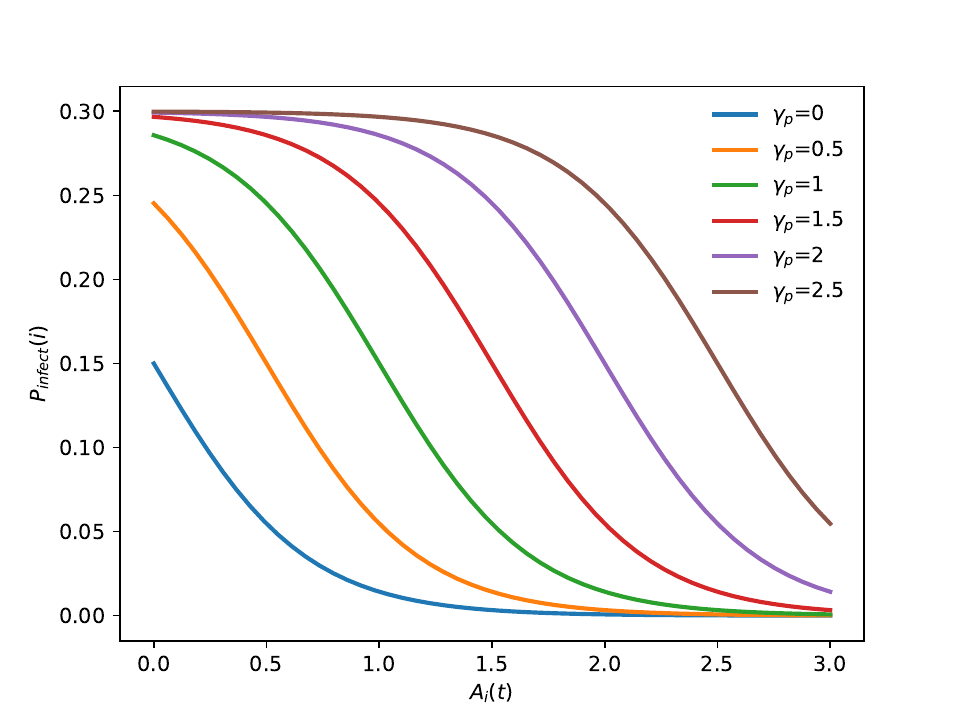}}
    \caption{\textbf{$P_{infect}(i)$ function diagram.} The figure illustrates the relationship between the infection probability $P_{infect}(i)$ and antibody level $A_i(t)$ for different parameter values. In figure (a), where $\gamma_p = 1.5$ and $\beta = 0.3$, the effect of $\alpha_p$ on the curve slope is shown. It can be observed from the different curves that as $\alpha_p$ increases, the curve tends more and more towards an S-shape. On the other hand, figure (b) shows the effect of $\gamma_p$ on the turning point of the curve for fixed values of $\alpha_p = 3$ and $\beta = 0.3$. As $\gamma_p$ increases, the curve shifts left and right, as depicted by the different lines. It should be noted that the antibody level $A_i(t)$ is restricted to the range of 0 to 3.}
\end{figure}
 Fig. 2 (a) displays the impact of $\theta$ and $\sigma$ on the antibody level $A_i(t)$. As per Eq. 2, $\sigma$ controls the magnitude of the Brownian motion in the change of the antibody level $A_i(t)$, thereby regulating the variability of the antibody level for each individual. On the other hand, $\theta$ controls the rate of change of $A_i(t)$, determining the rate of decay of the antibody from the time it is obtained. Fig. 2(a) shows that as $\sigma$ increases, the magnitude of the deviation of the antibody $A_i(t)$ from the mean value becomes larger. Additionally, the mean value of the antibody $A_i(t)$ decreases as $\theta$ increases, which is in line with the intuitive understanding of Eq. 2. Fig.2(b) illustrates the influence of $\gamma_p$ and $\alpha_p$ on the probability of infection $P_{infect}(i)$ for individual $i$. According to Equation 3, $\alpha_p$ is a slope parameter that governs the degree of influence of the antibody level on the probability of infection, while $\gamma_p$ is the parameter that controls the impact of the antibody level on the probability of infection. To provide a more intuitive representation, Fig. 2(b) plots the corresponding probability of infection under different parameters when the antibody level $A_i(t)=2$ and the base probability of infection $\beta=0.3$. Furthermore, we depict the folding line graph of infection probability under different parameters in Fig. 3.

The graphical representation in Fig. 3 depicts the variation of infection probability with $\alpha_p$ and $\gamma_p$ for different parameters. To ensure the model's accuracy and practicality, we commonly select $\alpha_p$ = 3 and $\gamma_p$ = 1.2 as optimal parameter values for numerical simulation experiments in the subsequent sections.

Next, we will focus on the time curves of the number of individuals in the three different states during the infection process. This curve will help us understand how the transmission occurs and assist us in analyzing whether the transmission has reached a steady state. We will conduct this numerical simulation within our constructed network, recording the values of the number of individuals in the S, I, R states at every time step from the first time step until the infection reaches a steady state, and then plot these curves.

We conducted simulated propagation experiments on WS small-world networks and BA scale-free networks. In these experiments, we used different antibody decay rates and chose networks with 1000 nodes for the simulations.

\begin{figure}[h]
    \centering
    \setlength{\abovecaptionskip}{-0.8cm} 
    \setlength{\abovecaptionskip}{-1.8cm} 
    \subfigcapskip=-2pt
    \subfigure[$\theta = 0.001$]{
        \includegraphics[width = 0.35\linewidth]{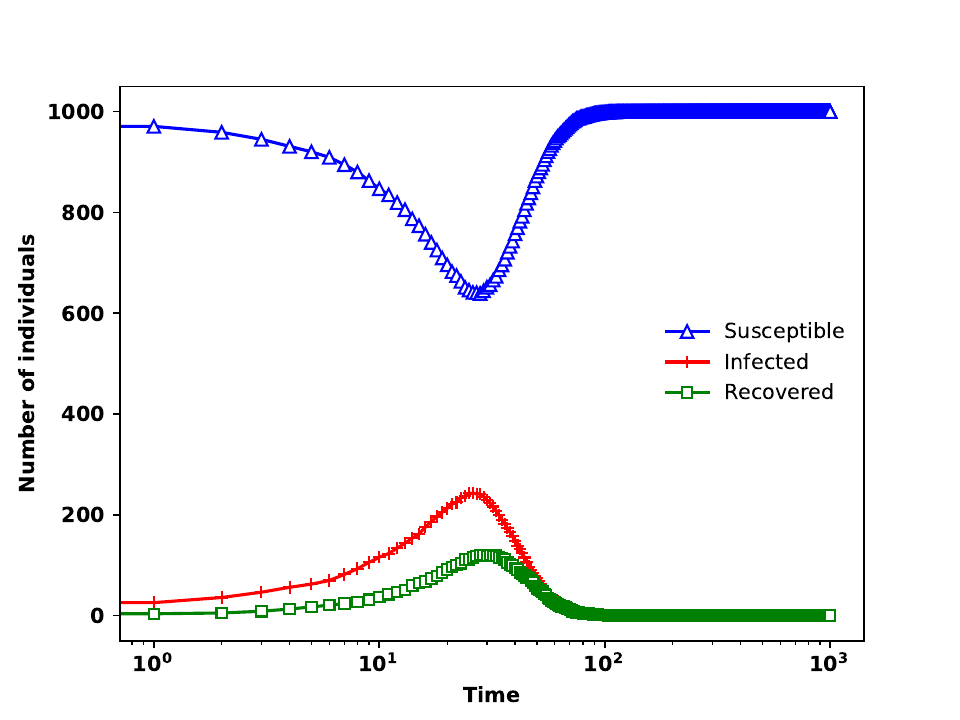}}
    \hspace{-0.8cm}
    \subfigure[$\theta = 0.005$]{
        \includegraphics[width = 0.35\linewidth]{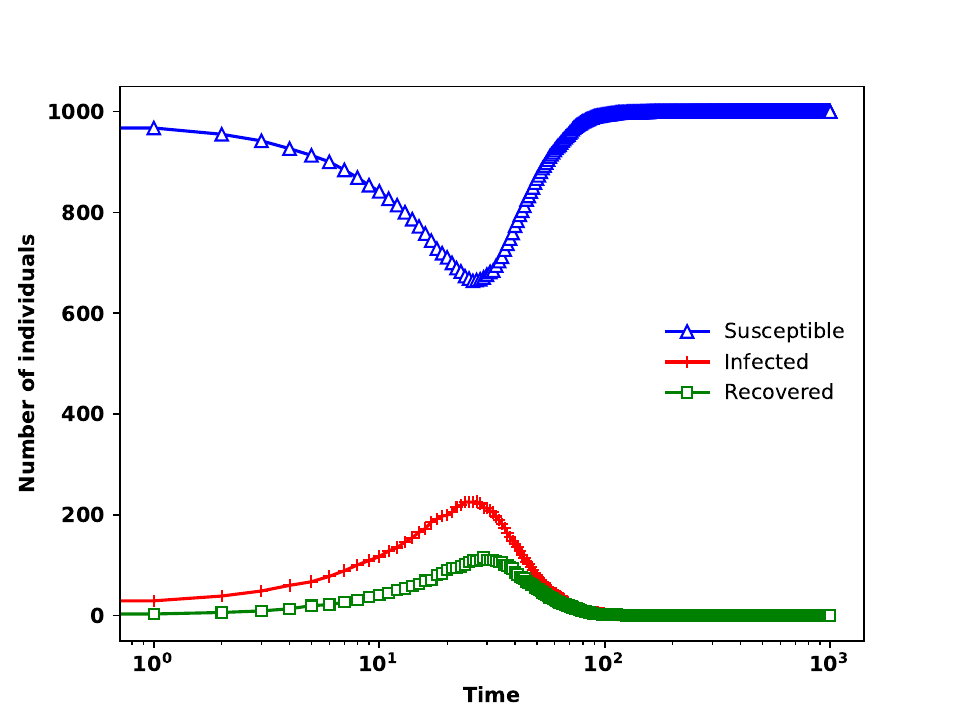}}
    \hspace{-0.8cm}
    \subfigure[$\theta = 0.01$]{
        \includegraphics[width = 0.35\linewidth]{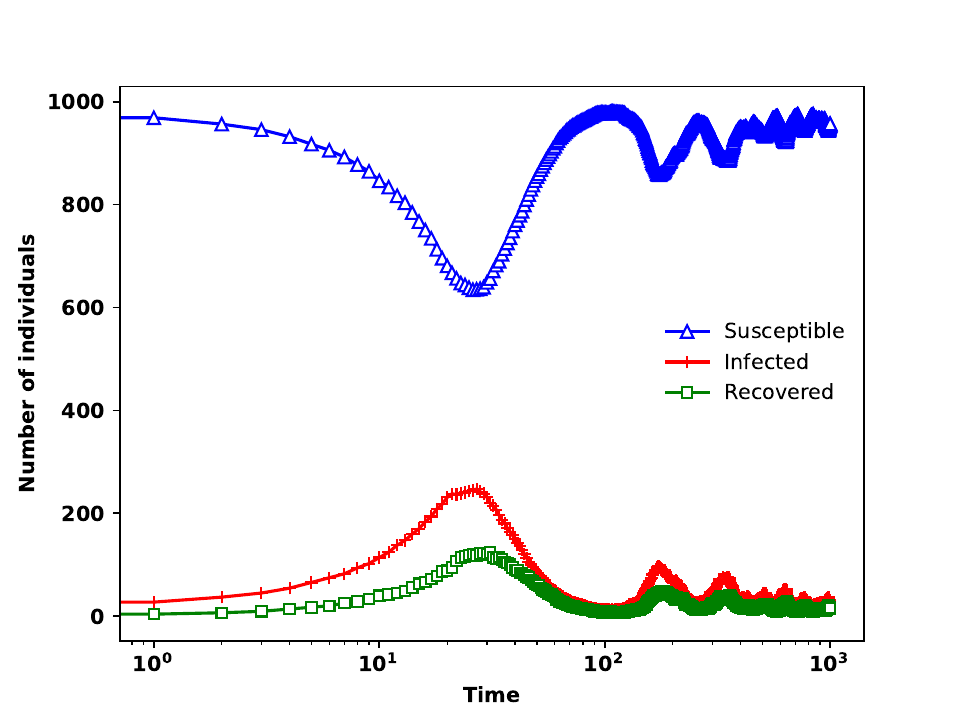}}
    \subfigure[$\theta = 0.02$]{
        \includegraphics[width = 0.35\linewidth]{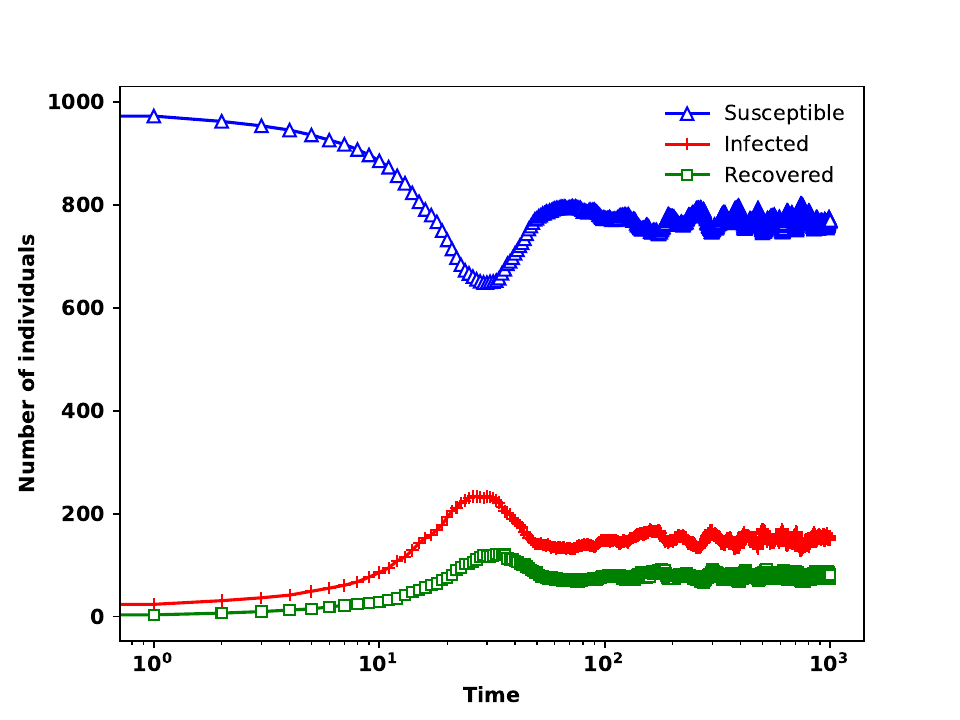}}
    \hspace{-0.8cm}
    \subfigure[$\theta = 0.05$]{
        \includegraphics[width = 0.35\linewidth]{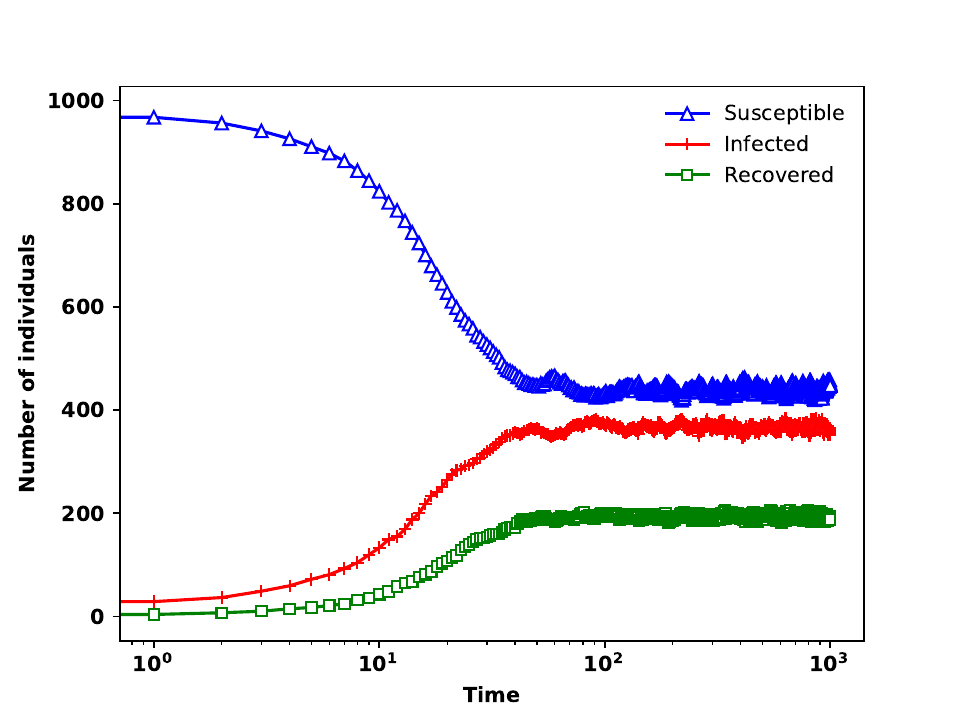}}
    \hspace{-0.8cm}
    \subfigure[$\theta = 0.1$]{
        \includegraphics[width = 0.35\linewidth]{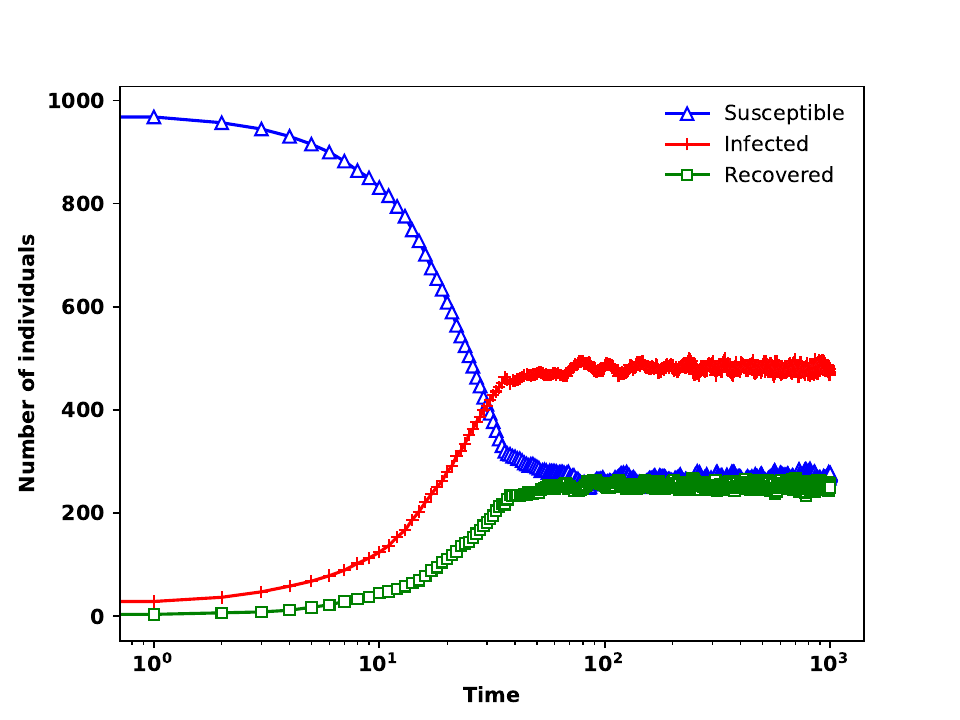}}
    \caption{\textbf{The propagation plots on the WS network.} It depicts the spread of the virus under different $\theta$ parameters. The experiments were conducted on a WS network with $N = 1000$, an average degree of $k = 4$, a disconnected edge reconnection probability of $p = 0.1$, a basic infection rate of $\beta = 0.2$, a disease recovery rate of $\mu = 0.1$, and regression values of antibody levels $\alpha = 0$. The horizontal axis of the plots shows the number of time steps in a logarithmic scale, while the vertical axis shows the number of individuals in each state during propagation, averaged over 50 iterations. As shown in plots (a), and (b), the disease eventually disappears, while in plots (c), (d), (e), and (f), the transmission reaches a plateau. These results suggest that the behavior of the virus is affected by the value of $\theta$.}
\end{figure}
The propagation process in the WS network is shown in Fig. 4. Based on the findings presented there, it can be concluded that the antibody decay rate $\theta$ plays a crucial role in determining the spread of the disease. The results demonstrate that as $\theta$ increases, the disease spreads more widely, infecting more people. When the antibody decay rate is very small, the disease dies out after one round of infection, as evidenced by the declining number of individuals in the infected state over time (Figs. 4(a)-(b)). However, with the increase in $\theta$, the disease goes through multiple rounds of infection before gradually dying out. The initial high antibody level of individuals in the network prevents further spread of the disease, but over time, the antibody level gradually decreases, leading to the peak of the second round of infection. The second round of the maximum number of infections is less than the first round, and after the second round of infections, the antibody levels of individuals get another boost. This process repeats, and eventually, the number of disease infections tends to 0. However, when $\theta$ is too large, the disease reaches a steady state during transmission, as the first round of infection has not yet been fully recovered, and the antibody level of the population drops enough to break out the next infection, as illustrated in Figs. 4(c)-(f). It is observed that the number of infected states in the steady state with different parameters increases with increasing $\theta$. These findings suggest that in WS networks, the size of $\theta$ significantly affects the propagation process. We also conducted propagation simulations on the BA network with the same parameters, and the results are presented in Fig. 5.

\begin{figure}[h]
    \centering
    \subfigcapskip=-2pt
    \setlength{\abovecaptionskip}{-0.8cm} 
    \setlength{\abovecaptionskip}{-1.8cm} 
    \subfigure[$\theta = 0.001$]{
        \includegraphics[width = 0.35\linewidth]{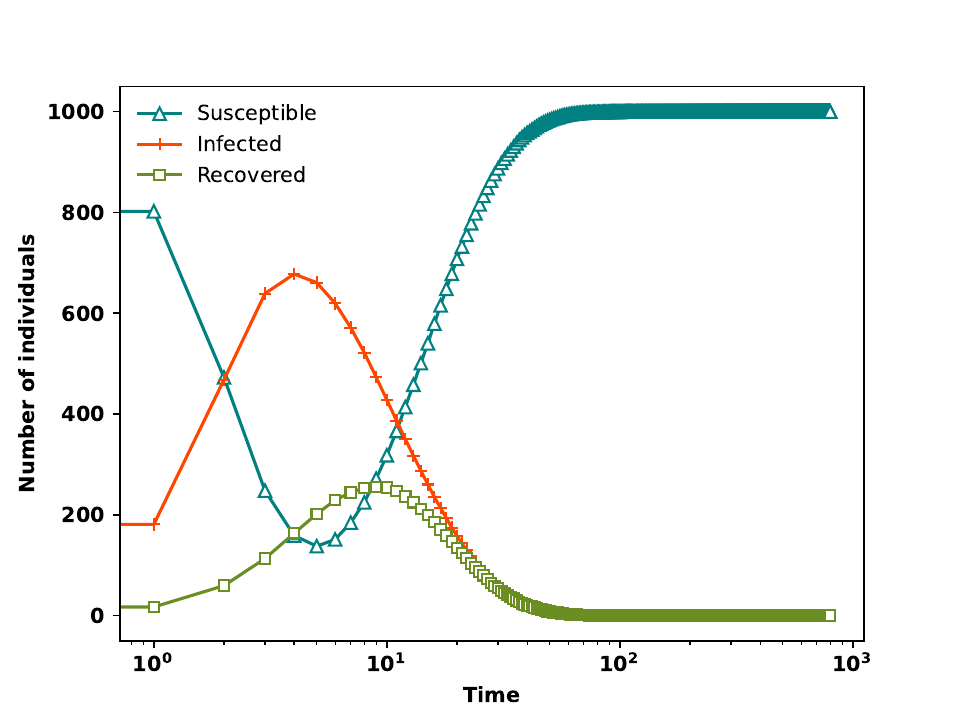}}
        \hspace{-0.8cm}
    \subfigure[$\theta = 0.005$]{
        \includegraphics[width = 0.35\linewidth]{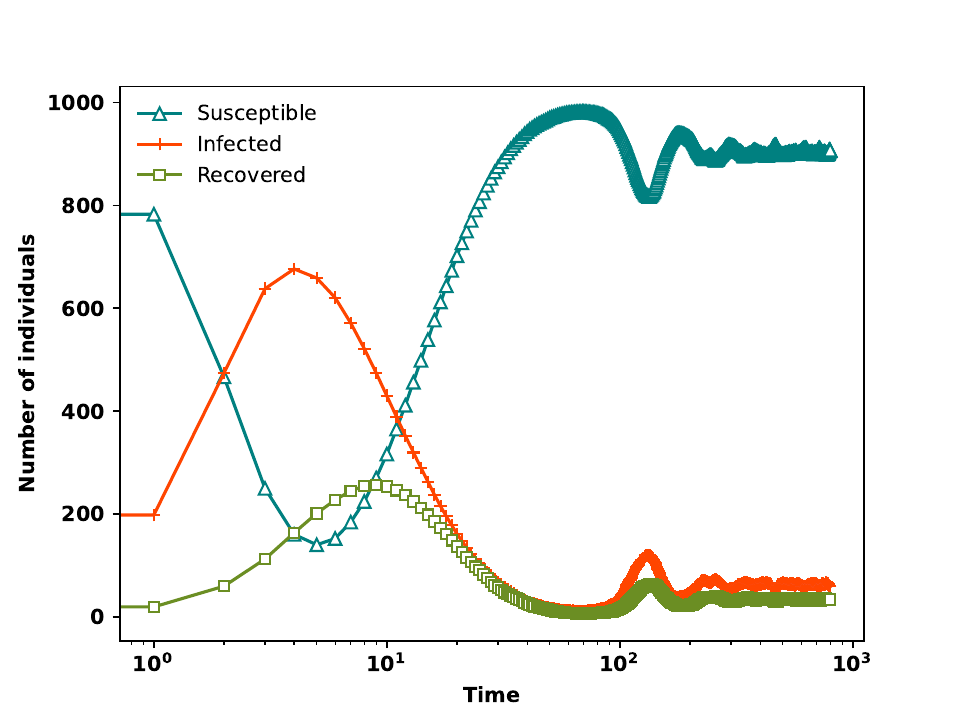}}
        \hspace{-0.8cm}
    \subfigure[$\theta = 0.01$]{
        \includegraphics[width = 0.35\linewidth]{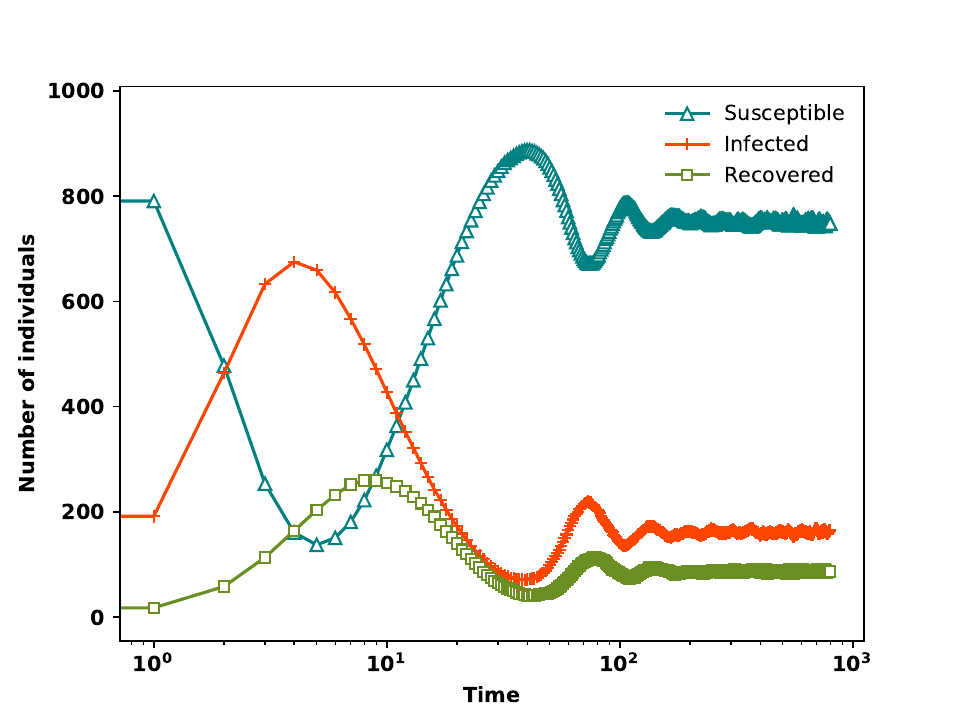}}
    \subfigure[$\theta = 0.02$]{
        \includegraphics[width = 0.35\linewidth]{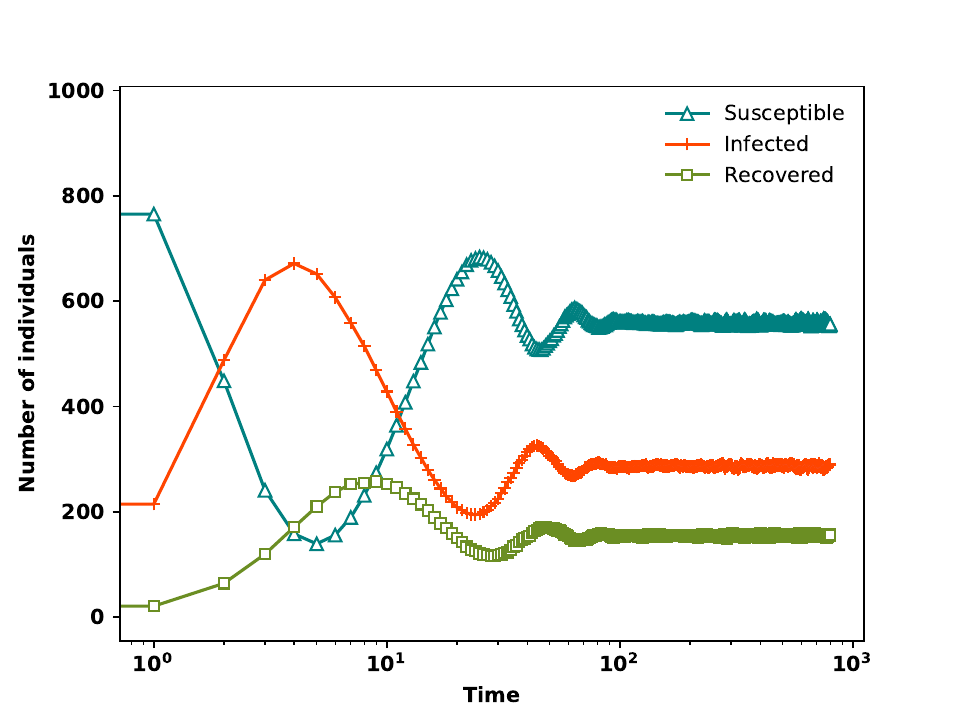}}
        \hspace{-0.8cm}
    \subfigure[$\theta = 0.05$]{
        \includegraphics[width = 0.35\linewidth]{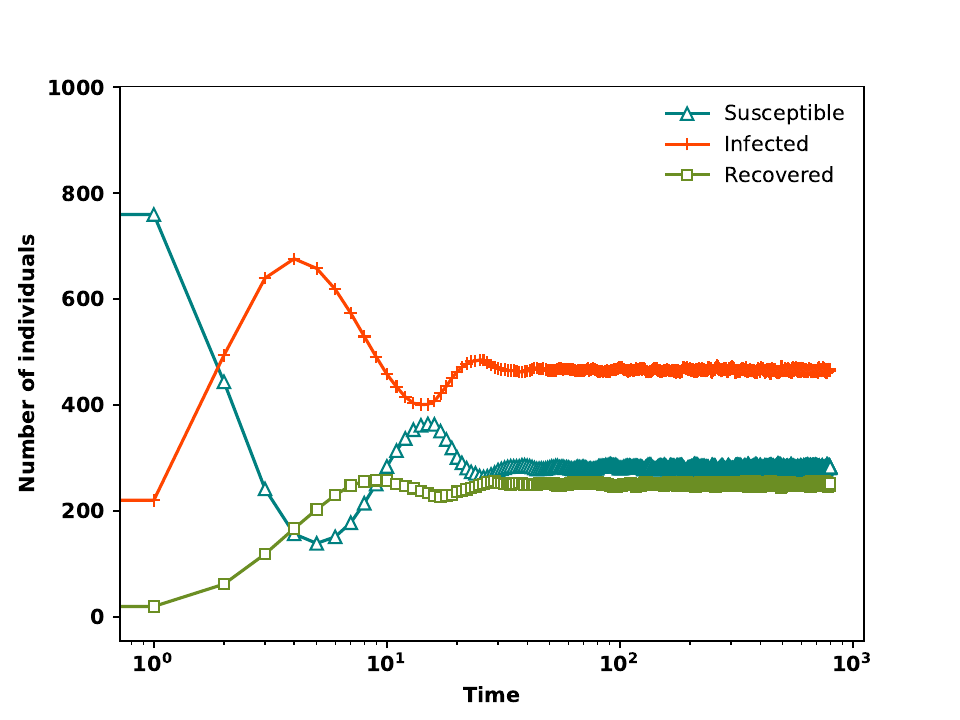}}
        \hspace{-0.8cm}
    \subfigure[$\theta = 0.1$]{
        \includegraphics[width = 0.35\linewidth]{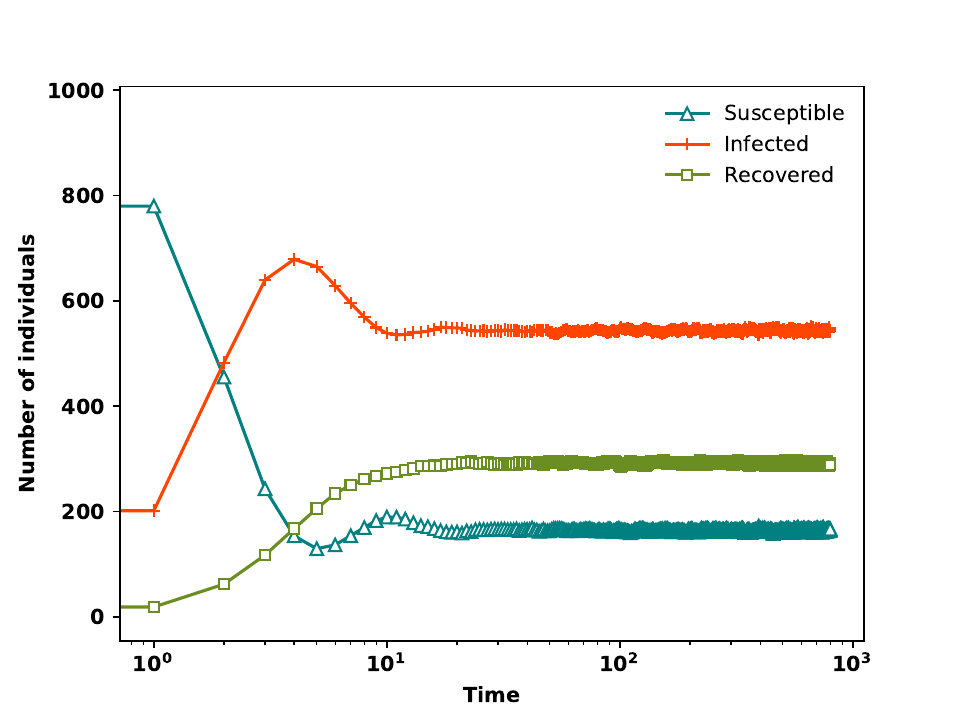}}
    \caption{\textbf{The propagation on a BA network.} where each subplot is labeled with its corresponding $\theta$ parameter value. The parameters used for the simulation are as follows: N = 1000 nodes in the BA network, with $k = 4$ as the network parameter, $\beta$ = 0.2 as the basic infection rate, $\mu$ = 0.1 as the disease recovery rate, and $\alpha$ = 0 as the regression values of antibody levels. The horizontal axis represents the number of time steps taken in a logarithmic scale, while the vertical axis indicates the number of individuals in each state during the propagation process. The data shown in the plot represents an average of 50 simulation runs. As observed, figure (a) shows that the disease eventually dies out, whereas figures (b)-(f) depict that the transmission eventually reaches a steady state.}
\end{figure}

The results presented in Fig. 5 reveal that the overall behavior of the BA network is not substantially different from that of the WS network. Specifically, Fig. 5(a) indicates that the propagation rate and scale of the BA network is greater than those of the WS network, as demonstrated by the comparison between Fig. 4(a) and Fig. 5(a). Furthermore, under the same parameters, while the disease eventually dies out on the WS network, it reaches a steady state on the BA network, suggesting that the propagation threshold of the BA network is more sensitive to the antibody decay rate. Additionally, as depicted in Fig. 4(c)-(f) and Fig. 5(c)-(f), the number of infected individuals after reaching the steady state is significantly higher in the BA network compared to the WS network. These observations suggest that the topology of the network plays a crucial role in the dynamics of disease propagation, and the findings obtained from simulations on the WS network can be generalized to other network models.

Finally, we will investigate the role and distribution of our newly proposed concept (antibodies) during transmission. This part will employ the same transmission process and network parameters as the previous two experiments. In this section, we can observe under what conditions antibodies will eventually reach zero and the entire process of their change. This will help us to further understand the realistic process of transmission.

We conducted an analysis of the stationary distribution of I-state individual at $\theta = 0.1$ for both network models. The results of this analysis are presented in Fig. 6.

\begin{figure}[h]
    \centering
    \subfigcapskip=2pt
    \subfigure[\textbf{Stationary distribution of I-state individuals on WS network}]{
        \includegraphics[width = 0.45\linewidth]{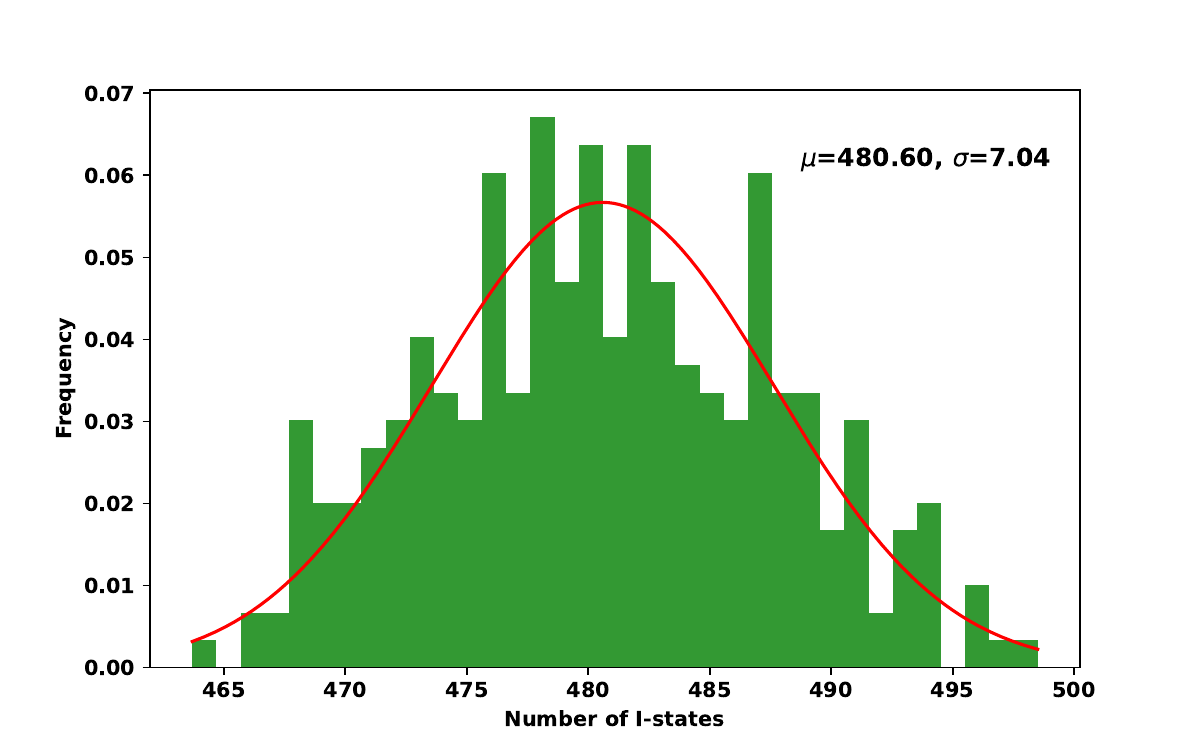}}
    \subfigure[\textbf{Stationary distribution of I-state individuals on BA network}]{
        \includegraphics[width = 0.45\linewidth]{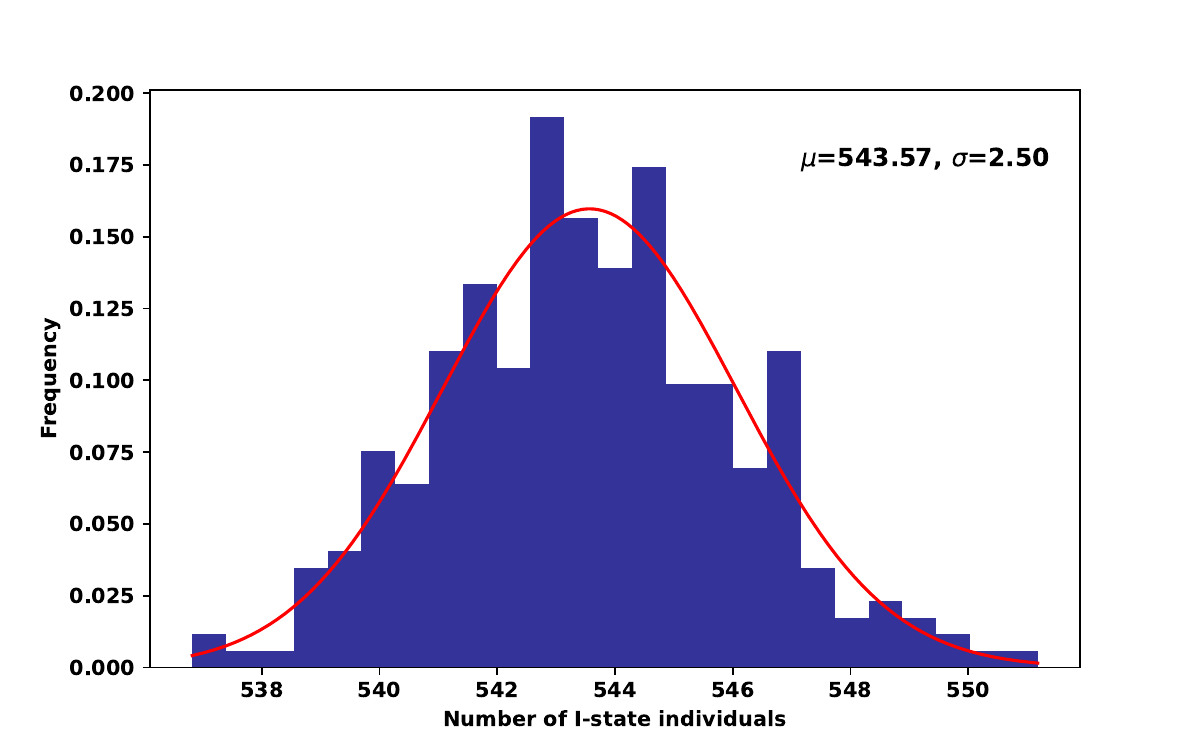}}
    \caption{\textbf{Stationary distribution of I-state individuals.} The parameters are the same as in Figs. 4 and 5 for the propagation process, we set $\theta$ to 0.1 and count the number of individuals in the I-state during the last 300 steps after the propagation has reached a steady state. We obtain a normal distribution for both (a) and (b), which we label with the mean $\mu_{Gauss}$ and standard deviation $\sigma_{Gauss}$ on the graph, and the red curve is the curve of normal distribution, and the green bar represents the probability of occurrence of each value. It can be observed from the graph that the mean on the WS network is smaller than on the BA network, while the standard deviation is larger.}
\end{figure}

We identified all values with frequencies greater than or equal to 2 and generated Fig. 6 to illustrate the results. Our findings reveal that the frequency distribution for both WS and BA networks conforms to a normal distribution, with $\mu_{Gauss}$ 480.6 and 543.57 and $\sigma_{Gauss}$ 7.04 and 2.5, respectively. These results suggest that the I-state numbers under the BA network will be more tightly clustered, while the I-state numbers under the WS network will be more widely dispersed.

Our experiments on the WS small-world network reveal that the model has a limited propagation range and relatively slow propagation speed, owing to the highly aggregated topology and short path characteristics of WS small-world networks. The closely connected nodes in the network make it difficult for information to reach the edges of the network. Additionally, we have observed that the antibody decay rate has a significant impact on the spread range and spread speed of the disease, with faster decay rates leading to a significant increase in both the spread range and speed of the disease. On the other hand, in the BA scale-free network, the model exhibits wider spread and faster propagation. This is attributed to the power-law distribution property of the network topology, where a few nodes with larger degrees become the key nodes for propagation.

Our experimental results highlight that network topology and antibody decay rate are key factors influencing the propagation of the model. Further, different network structures exhibit significant differences in the effect on the propagation of this model.

Figs. 4 - 6 illustrate the changes in the number of individuals in each state throughout the transmission process, with changes in the probability of infection attributed to variations in antibody levels. To gain a more thorough understanding of the transmission process, it is imperative to conduct a more in-depth examination of the fluctuations in antibody levels. Therefore, we have generated Fig. 7, which displays the curve of mean antibody levels for the entire network throughout the transmission process.

\begin{figure}[h]
    \centering
    \subfigcapskip=2pt
    \subfigure[\textbf{WS network propagation process antibody change}]{
        \includegraphics[width = 0.45\linewidth]{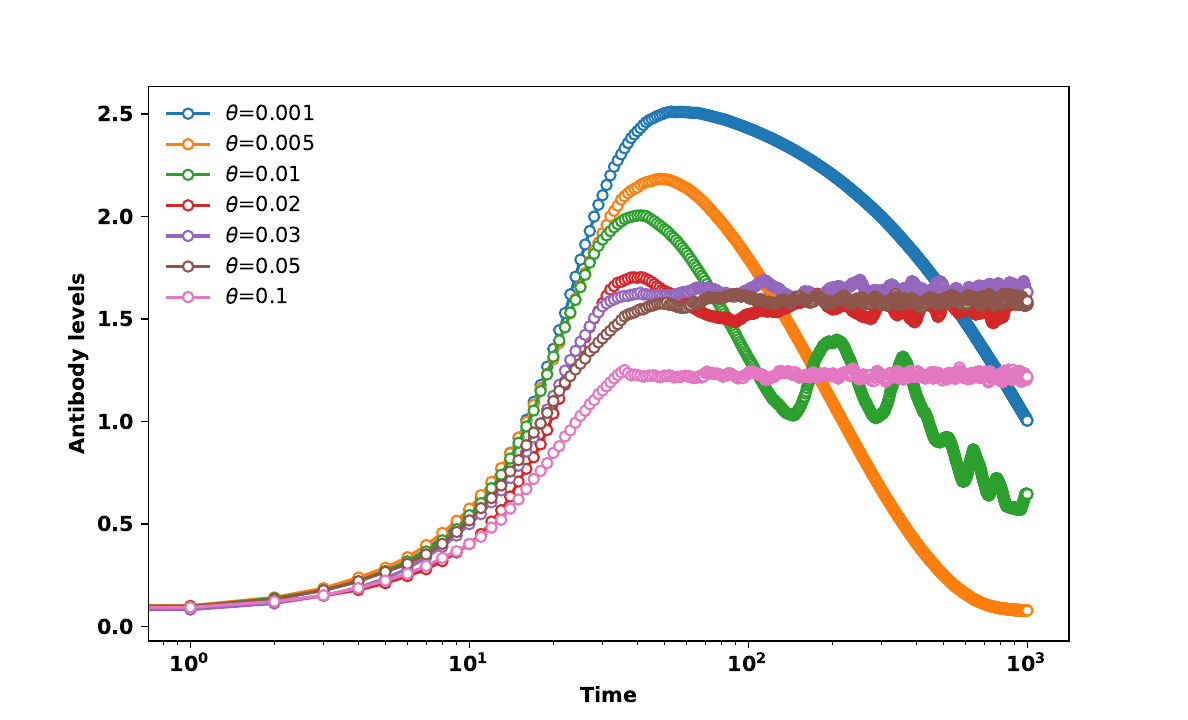}}
    \subfigure[\textbf{BA network propagation process antibody change}]{
        \includegraphics[width = 0.45\linewidth]{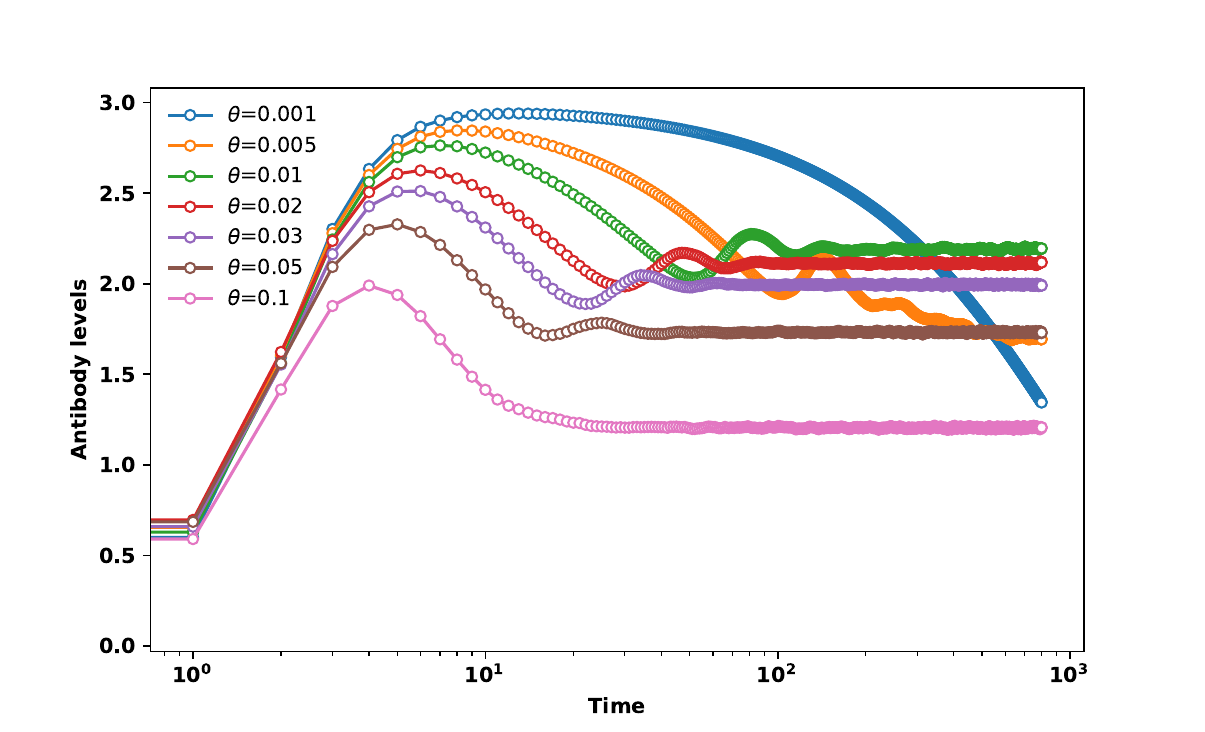}}
    \caption{\textbf{Variation of the mean value of antibody.} The parameters used for the (a)(b) process correspond to the transmission processes depicted in Figs. 4 and 5, respectively. The different colored lines in Fig. 7 represent the changes in the mean value of antibody levels for different $\theta$ parameters, which correspond to the propagation processes in Figs. 4 and 5. It is evident that after it peak, some curves gradually converge to 0, while others attain a steady state. }
\end{figure}

When examining Fig. 7, it can be inferred that the antibodies will eventually converge around a specific value. There exists a certain interval of values for $\theta$ where antibodies will eventually return to zero, while for the interval where antibodies will not return to zero, it is observed that the higher the value of $\theta$, the smaller the eventual steady state value of the antibodies. When analyzing the peak state of the initial round of infection, it is found that lower $\theta$ values result in higher peak values of the mean antibody, indicating that $\theta$ has a negative effect on the peak range of the antibody. For the aspect of time to peak, there is no difference between different $\theta$ values.

Similar to Fig. 6, we analyzed the stationary distribution of antibody levels on both networks at $\theta = 0.1$ and the results are shown in Fig. 8.

\begin{figure}[h]
    \centering
    \subfigcapskip=2pt
    \subfigure[\textbf{Stationary distribution of antibody levels on WS network}]{
        \includegraphics[width = 0.45\linewidth]{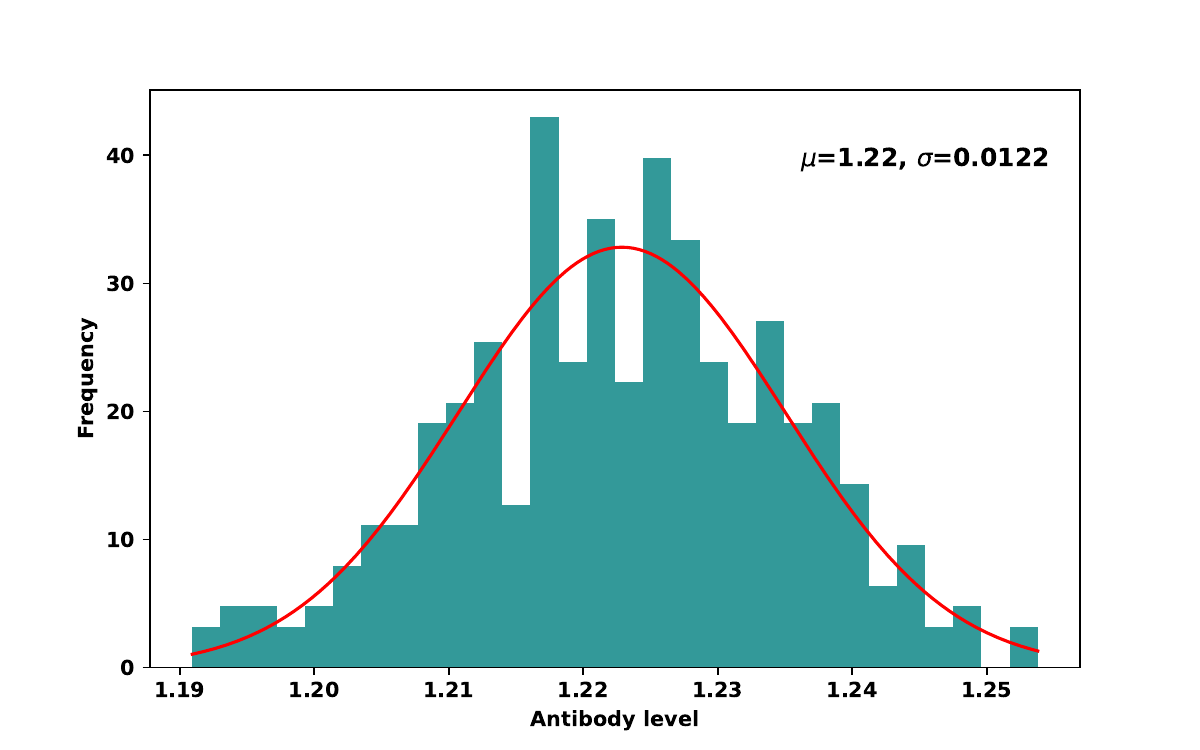}}
    \subfigure[\textbf{Stationary distribution of antibody levels on BA network}]{
        \includegraphics[width = 0.45\linewidth]{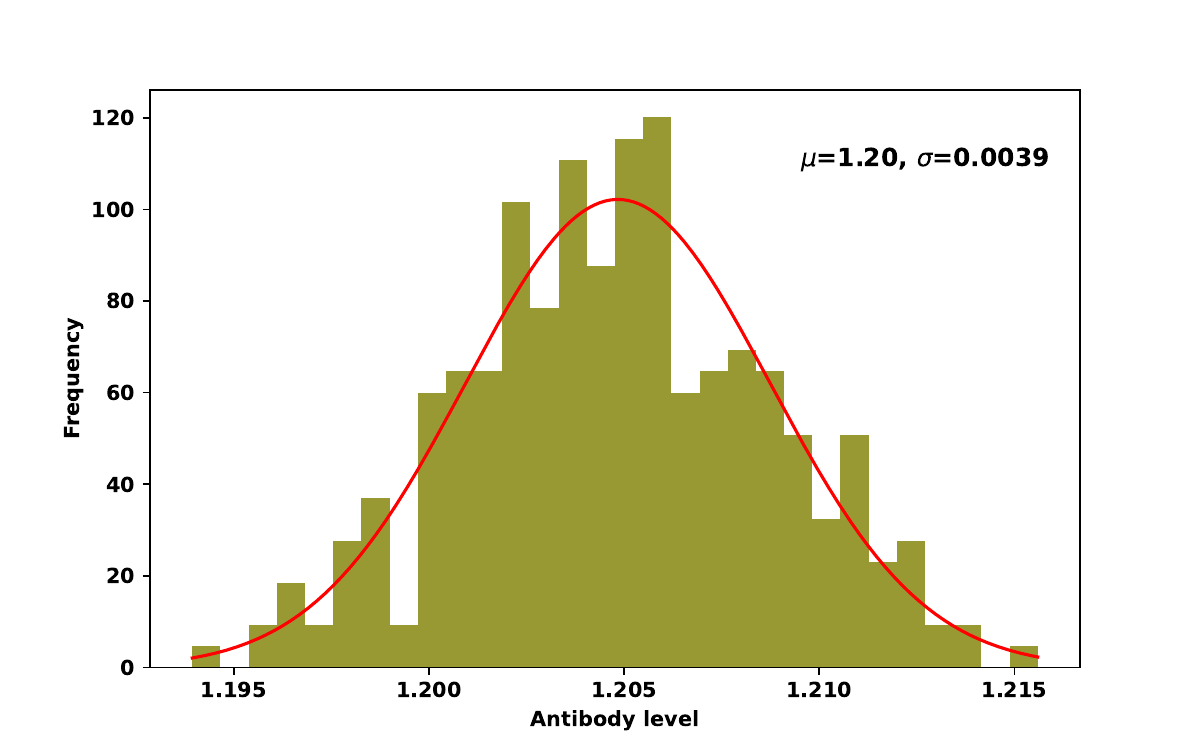}}
    \caption{\textbf{Stationary distribution of antibody levels.} The parameters of the propagation process are the same as Fig. 4 and Fig. 5, and we take the mean value of the antibody level in the last 300 steps after the statistical propagation stable when $\theta = 0.1$, and finally we get the stationary distribution (a) (b) are the normal distribution, and we also label the mean $\mu_{Gauss}$ and standard deviation $\sigma_{Gauss}$ of the normal distribution in the figure, where the red curve is the curve of the normal distribution, respectively. The blue bars represent the probability of occurrence of each value. From the figure, it can be seen that the mean values on the two networks do not differ much, while the standard deviation of the stationary distribution of antibodies on the WS network is significantly larger than that on the BA network.}
\end{figure}

Observing Fig. 8, the stationary distribution of the final antibody levels on both networks is close to a normal distribution, with $\mu_{Gauss} = 1.22$ and $\sigma_{Gauss} = 0.0122$ under the WS network and mean $\mu_{Gauss} = 1.2$ and standard deviation $\sigma_{Gauss} = 0.0039$ under the BA network. Consistent with the conclusions obtained previously, antibody levels are again more aggregated under the BA network and more dispersed under the WS network. However, the difference between the two antibody level means is not significant, and in combination with Fig. 7, it can be seen that probably at larger values of $\theta$, it is the most dominant among the factors influencing antibody levels.

To further explore the $\theta$ range of values that will eventually return the antibody to zero, we conducted another experiment investigating the influence of the $\sigma$ and $\theta$ range on the eventual smooth value of the antibody level. The experimental outcomes are illustrated in Fig. 9.

\begin{figure}[h]
    \centering
    \subfigcapskip=2pt
    \subfigure[\textbf{Antibody stable value in WS network}]{
        \includegraphics[width = 0.49\linewidth]{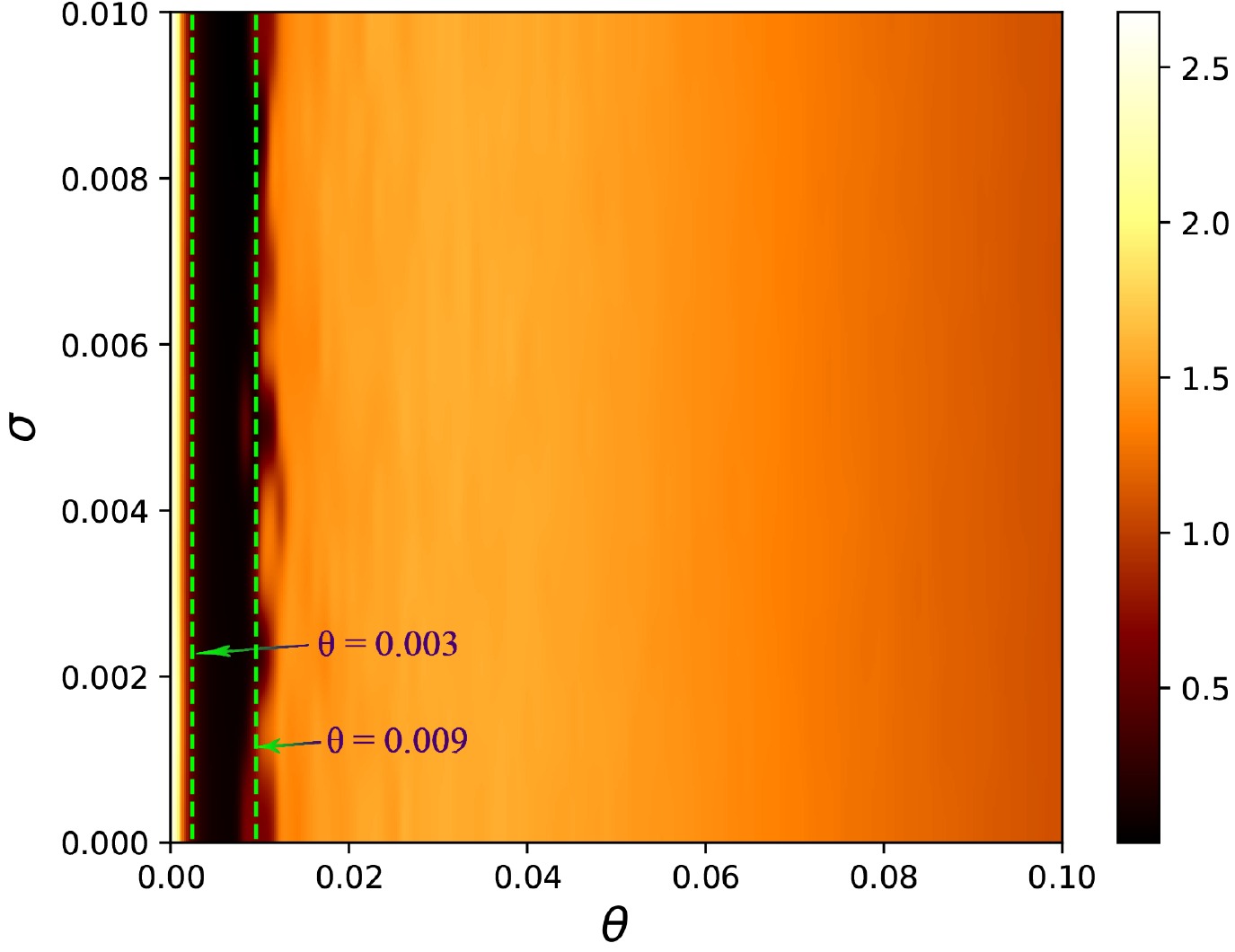}}
    \subfigure[\textbf{Antibody stable value in BA network}]{
        \includegraphics[width = 0.49\linewidth]{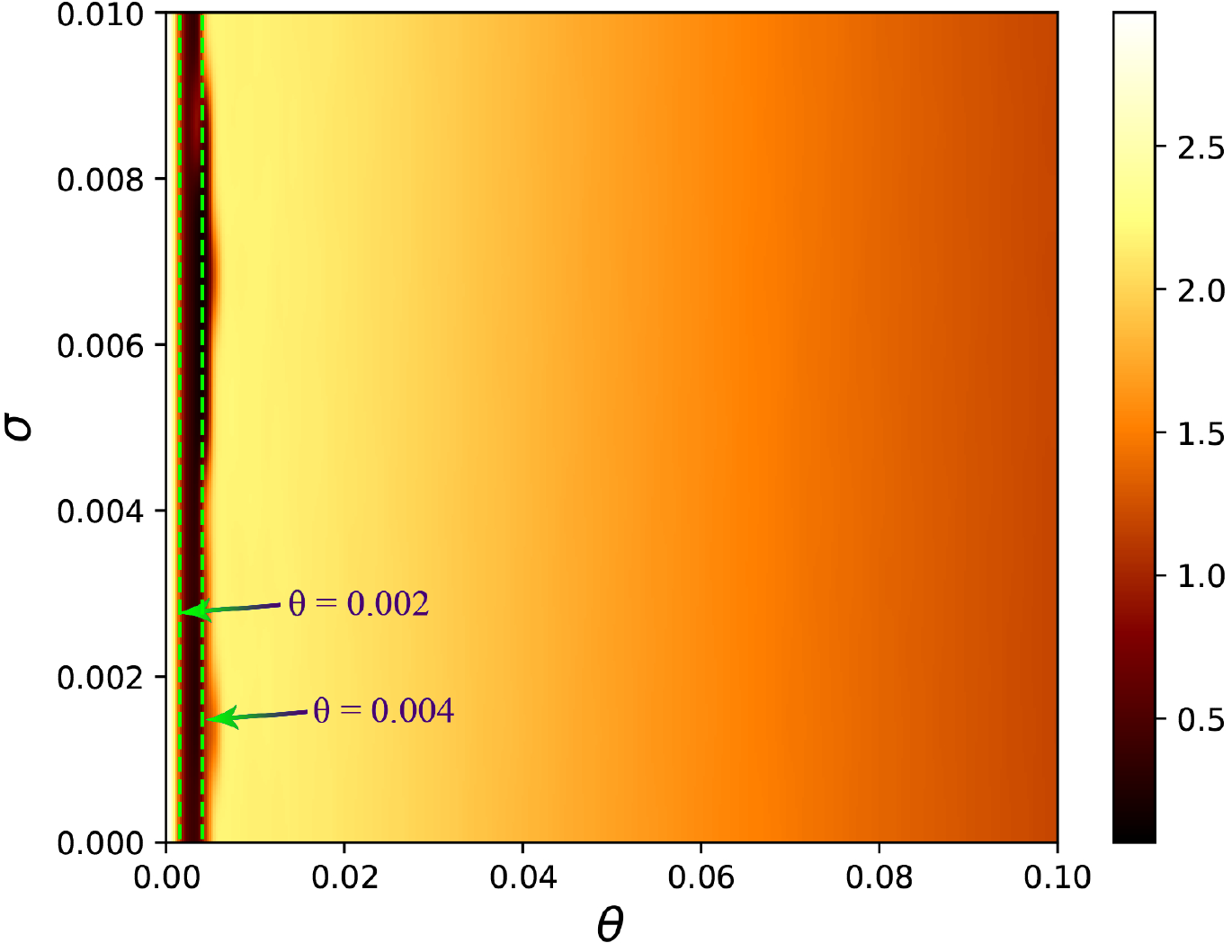}}
    \caption{\textbf{Antibody stable value.} The experimental results shown in (a) and (b) investigate the effect of different values of $\theta$ and $\sigma$ on the final antibody level. The horizontal axis represents the values of $\theta$, while the vertical axis represents the values of $\sigma$. In (a), the experiment is conducted on the ws network with parameters of $N=1000$, average degree $k=4$, broken edge reconnection probability $p=0.1$, basic infection rate $\beta=0.2$, disease recovery rate $\gamma=0.1$, and regression values of antibody levels $\alpha=0$. In (b), the experiment is conducted on the BA network with the network construction parameter of $k=4$ and the rest of the parameters being the same as those of the BA network. The final mean antibody level is taken as the average after propagation for a sufficiently long time and the last 50 steps of the antibody are taken for averaging. The graphs clearly show that there exists a specific range of $\theta$ values that cause the mean final antibody level to converge to 0.}
\end{figure}

Fig.9 clearly indicates that $\theta$ has a significant impact on the final antibody level, while the impact of $\sigma$ can be disregarded. This is because $\sigma$ is a parameter that represents the intensity of the Brownian motion, and as the number of iterations increases, the effects of random fluctuations decrease, resulting in a gradual convergence of $\sigma$ to its mean value. This convergence may lead to a stabilization of the system, which could explain why the impact of $\sigma$ on the final antibody level can be ignored. Therefore, we focus on the impact of $\theta$. The figure shows that there exists a specific range of $\theta$ values that results in the average final antibody level converging to 0. In Fig. 9, we label this range as [0.003, 0.009] for the WS network and [0.002, 0.004] for the BA network. When $\theta$ is within this range, the disease ultimately dies out. However, if $\theta$ is taken slightly larger than the right boundary of the range, the final antibody level rapidly increases to a value greater than 0, and as $\theta$ continues to increase, the final antibody level decreases toward 0. This observation is reflected in the graph as the color on the right side of the black bar line rapidly lightening, followed by a gradual darkening from left to right.

It should be noted that when the antibody decay rate $\theta$ is very low (less than the value indicated on the right side of the above figure), the virus will disappear from the network. However, within a finite number of time steps, due to the small decay rate of antibodies ($\theta$), the network's average antibody level will not revert to zero by the time of our statistics. This part of the simulation experiment might explain the transmission of certain viruses that provide lifetime immunity after a single infection, such as rubella and HFMD. Additionally, our SIRS model degenerates into an SIR model with antibodies under this specific condition.

In summary, our simulation experiments on the WS small-world network and the BA scale-free network have revealed that the $\theta$ value plays a critical role in determining the final steady state of disease transmission and antibody levels, while the effect of the $\sigma$ value is negligible. Notably, the width of the bar line indicating the range of $\theta$ values leading to a converging final antibody level is smaller for the BA network compared to the WS network. This observation is consistent with the BA network having a higher propagation speed and a greater sensitivity to the decay rate of the antibody. These findings provide valuable insights into the underlying mechanism of disease transmission and antibody production, and may help to guide the development of effective disease control strategies.

\section{Conclusion and outlook}

In conclusion, we presents a novel mathematical model that integrates the presence of an antibody retention rate to investigate infection patterns of individuals who have survived multiple infections. The model employs a system of stochastic differential equations to derive the equilibrium point, threshold and provides rich experimental results through numerical simulations to further elucidate the propagation properties of the model. The findings offer valuable insights for epidemic prevention and control in practical applications. Specifically, this study highlights that network topology and antibody decay rate are key factors that significantly influence the propagation of the model.

Our model also has some limitations. First, the process describing antibody dynamics may require further investigation; the Ornstein-Uhlenbeck (OU) process is a very simple description, and the actual situation is more complex. Second, the numerical simulation may not be comprehensive enough and needs to be further explored in future work.

This study lays the groundwork for future development and refinement of the model to provide more accurate predictions and insights into the spread of epidemics. Moreover, the findings can inform and improve epidemic prevention and control strategies. Future research directions could extend the model to incorporate additional factors that may influence the transmission of infectious diseases. Additionally, the applicability of the model to other types of networks and real-world scenarios is worth exploring.e

\section*{Acknowledgment}
This work was supported by the National Natural Science Foundation of China (NSFC) (Grant No.62206230) and the Natural Science Foundation of Chongqing (Grant No. CSTB2023NSCQ-MSX0064).

\section*{Declaration of Competing Interest}
The authors declare that they have no known competing financial interests or personal relationships that could have appeared to influence the work reported in this paper.

\section*{Data Availability}
For inquiries regarding data availability, please contact the corresponding author, Minyu Feng, at myfeng@swu.edu.cn.

\bibliographystyle{unsrt}
\bibliography{ref}

\end{document}